\documentclass[prl,aps,twocolumn,superscriptaddress]{revtex4-1}
\usepackage{graphicx,color}
\usepackage{amsthm}
\usepackage{amsfonts}
\usepackage{algorithmic}
\usepackage{enumerate}
\usepackage{latexsym}
\usepackage{amsmath}
\usepackage{amssymb}
\usepackage[colorlinks=true,citecolor=blue,linkcolor=blue]{hyperref}

\emergencystretch=\maxdimen
\begin{document}

\title{Different phase leads to different transport behavior in Pb$_9$Cu(PO$_4$)$_6$O compounds}

\author{Ran Liu}
\affiliation{Department of Physics, Beijing Normal University, Beijing 100875, China\\}
\author{Ting Guo}
\affiliation{Department of Physics, Beijing Normal University, Beijing 100875, China\\}
\author{Jiajun Lu}
\affiliation{School of Physics and Electronics, Shandong Normal University, Jinan, 250358, China\\}
\author{Junfeng Ren}
\affiliation{School of Physics and Electronics, Shandong Normal University, Jinan, 250358, China\\}
\author{Tianxing Ma}
\email{txma@bnu.edu.cn}
\affiliation{Department of Physics, Beijing Normal University, Beijing 100875, China\\}
\affiliation{Key Laboratory of Multiscale Spin Physics, Ministry of Education, Beijing 100875, China\\}

\begin{abstract}
The recent claimed room-temperature superconductivity in Cu-doped lead apatite at ambient pressure are under highly debate.
To identify its physical origin, we studied the crystal structures, energy band structures, lattice dynamics and magnetic properties of the parent Pb$_{10}$(PO$_4$)$_6$O compound, in which two different phases of the LK-99 compound are analyzed in detail.
Our results show that the Pb$_{10}$(PO$_4$)$_6$O compound is an indirect band gap semiconductor, where Cu doping at the 4$f$ site of Pb leads to a semiconducting to half-metallic transition. Two half-filled flat bands spanning the Fermi energy levels are present in the 4$f$-phase of LK-99, which are mainly formed by hybridization of the $d_{x^2-y^2}$ and $d_{zy}$ orbitals of Cu with the 2$p$ orbitals of O. In addition, 6$h$-phase of LK-99 always has spin polarity at the bottom of the conduction band and at the top of the valence band, making the material a bipolar magnetic semiconductor. Our results are basically consistent with the recent experimental transport properties of LK-99 posted on arXiv:2308.05778.
\end{abstract}

\noindent


\maketitle

\noindent
{\it{Introduction}}
Since the discovery of superconductivity in 1911\cite{1911The},
extensive studies have been performed more than one hundred a year on how to improve the critical temperature for the appearance of superconductivity. In past century, the superconducting transition temperature of various types of superconductors has achieved a rapid improvement\cite{PhysRevLett.58.908,1986ZPhyB..64..189B,Zhao2016,85fbf29a80c2455db387a987f4d06336}, from 4.2 K in metal Hg\cite{1911The} to 133 K in HgBa$_2$Ca$_2$Cu$_3$O$_{1+x}$\cite{Schilling1993} under ambient pressure.
Under high pressure, the maximum superconducting transition temperature has reached 262 K in yttrium superhydride\cite{PhysRevLett.126.117003}, 203 K in hydrogen sulfide materials\cite{Drozdov2015}, as well as 250 Kelvin in LaH$_{10}$\cite{Drozdov2019}.
Despite these developments, the superconducting transition temperature of most superconductors under ambient pressure is far bellow the room temperature, and the high pressure required for high-temperature superconductors are significant obstacles to promote superconducting applications\cite{Sun2023}.
Room temperature superconductors under ambient pressure remains as mysterious, and its realization survives as Holy Grail.

Recently, a highly anticipated study claimed that room temperature superconductivity has been achieved in copper doped lead apatite materials under ambient pressure\cite{lee2023roomtemperature,lee2023superconductor}, and this sample also was called as LK$\text{-}$99.
LK$\text{-}$99 material is mainly prepared by mixing four materials: PbO, PbSO$_4$, Cu, and P. Its chemical formula is Pb$_{10-x}$Cu$_x$ (PO$_4$)$_6$O ($0.9<x<1.1$). According to the definition of the wyckoff position of the crystal, the sites of Pb ions can be divided into $4f$ and $6h$\cite{zhang2023structural}.
It was pointed out that when the doped copper ions replaced the 4$f$ Pb ions, which caused distortion in the crystal structure with a volume reduction of $0.48\%$, leads to a slight shrinkage of lattice constant, and superconducting quantum states should be generated. It was argued that the key to the superconductivity of LK$\text{-}$99 at room temperature and ambient pressure lies in its ability to maintain the structural distortion caused by copper doping\cite{lee2023roomtemperature}. These experimental reports, provoke scholarly excitement\cite{griffin2023origin,cabezasescares2023theoretical,zhang2023structural,tao2023cu,si2023pb10xcuxpo46o,krivovichev2023crystal,yue2023correlated,wu2023observation,bai2023ferromagnetic,
jiang2023pb9cupo46oh2,jain2023phase,mao2023wannier,hao2023firstprinciples,hou2023current,chen2023statistical,singh2023experimental,singh2023experimental,panas2023entertaining,witt2023superconductivity,georgescu2023cudoped}.

A thorough study of the electronic structure of this material is urgent to unraveling the mystery of LK$\text{-}$99. By using density functional theory (DFT), Escares and Barrera found that Cu ions seems to be hosts in the lattice and a nearly flat band emerge around the Fermi level\cite{cabezasescares2023theoretical}. Griffin claimed that there were isolated flat bands around the Fermi level in LK$\text{-}$99 material, and he believed that the origin of these isolated flat bands is the structural distortion caused by Cu ions substitution of $4f$ Pb ions, which may result in a higher transition temperature for LK$\text{-}$99 compared to existing high-temperature superconductors\cite{griffin2023origin}. Zhang et al. argued that Cu ions substitute not $4f$ Pb ions but $6h$ Pb ions, which have lower energy than the former, and if $6h$ Pb ions were substituted by Cu ions, the LK$\text{-}$99 material should be a semiconductor with an energy gap of 0.9 eV in its ground state\cite{zhang2023structural}.
Form these viewpoints, different ways of substitution would result in completely different transport properties in LK$\text{-}$99 material. The horizontal comparison of the conductivity, lattice structure, electronic structure, and other properties of two different phases of LK$\text{-}$99 should be a fundmental step in understanding the novel physical properties of this material.

\noindent
{\it Methods}
Our DFT calculations were performed by using the QUANTUM ESPRESSO (QE) software package\cite{Giannozzi_2009,Giannozzi_2017,10.1063/5.0005082}. We adopted the generalized gradient approximation (GGA) of the Perdew-Burke-Ernzerhof (PBE)functional for the exchange-correlation energy\cite{PhysRevLett.77.3865}, and the projector augmented-wave (PAW) pseudopotentials were employed to describe the electron-ion interaction\cite{PhysRevB.59.1758}. After test of complete convergence, the kinetic energy cutoffs of wave functions and charge densities cutoffs were set to $50 Ry$ and $350 Ry$, respectively. The charge density was calculated self-consistently on an unshifted mesh of $4\times4\times6$ $k$-points and the Methfessel-Paxton broadening of $0.005 Ry$ was adopted \cite{PhysRevB.40.3616}.

In the optimization of geometry structural, the convergence thresholds for total energy and ionic force were set to $10^{-4} Ry$ and $10^{-3} Ry/Bohr$, respectively. All geometric structures were fully optimized in our calculations to achieve the minimum energy. Considering the presence of unpaired isolated electrons on $3d$ orbital of Cu$^{2+}$, the spin polarization in all calculations was switched on when calculating the properties of Pb$_9$Cu(PO$_4$)$_6$O compound and was contrastively switched off in Pb$_{10}$(PO$_4$)$_6$O compound. Density functional perturbation theory (DFPT) was used to calculate the phonon dispersion spectrum. The dynamics matrix and perturbation potential were computed on a mesh of $2\times2\times2$ $q$-points centered at the $\Gamma$ point\cite{RevModPhys.73.515}.

\begin{figure}[tbp]
\includegraphics[scale=0.05]{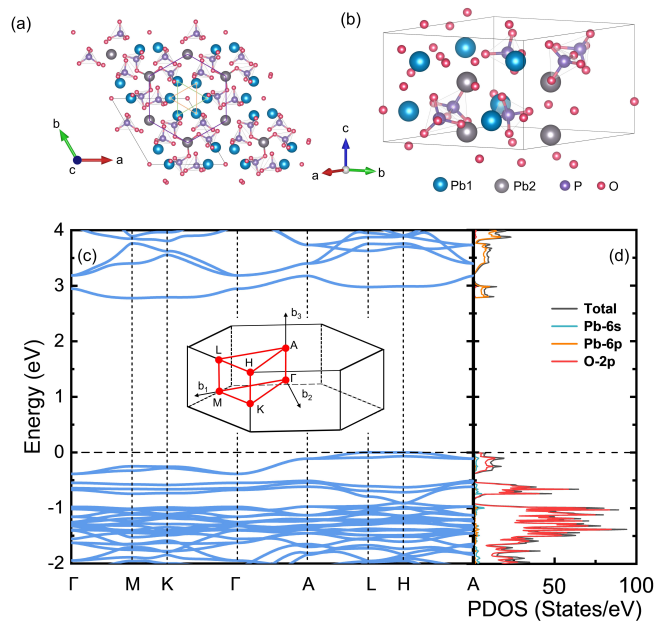}
\caption{(a) Top view of the crystal structure of Pb$_{10}$(PO$_4$)$_6$O compound. The black line denotes the unit cell. Pb1 ions form two triangles of different layers with opposite shapes, marked by green and orange lines, respectively. The Pb2 ions form a hexagon marked by purple lines. (b) Side view of the crystal structure of Pb$_{10}$(PO$_4$)$_6$O compound. (c) The band structure of Pb$_{10}$(PO$_4$)$_6$O compound. The Fermi energy level is set to zero. The inset shows the high symmetry k-point selected for plotting the band structure. (d) The Orbital-resolved DOS of Pb$_{10}$(PO$_4$)$_6$O compound.
} \label{crystal structure1}
\end{figure}

\noindent
{\it Results and discussion} Fig.\ref{crystal structure1}(a) and Fig.\ref{crystal structure1}(b) illustrate the crystal structure of lead-apatite, which serves as the parent compound of LK$\text{-}$99, with a chemical formula of Pb$_{10}$(PO$_4$)$_6$O. After performing a full structural optimization, the lattice constants were determined to be $a=10.044 \AA$, $c=7.502 \AA$ and the volume of a unit cell is $654.8 \AA^3$, which are consistent well with the experimental values of $a=9.865 \AA$, $c=7.431 \AA$ and the volume of a unit cell is $626.3 \AA^3$\cite{KrivovichevBurns+2003+357+365}. It should be noted that the predicted larger lattice parameters are a result of the overestimation of the employed PBE functional. Within the lead-apatite structure, there are two symmetry inequivalent Pb ions, namely Pb1 and Pb2. The Pb1 ions form two triangles of different layers with opposite shapes, while the Pb2 ions arrange themselves in a hexagonal fashion, as that depicted in Fig.\ref{crystal structure1}(a). According to the definition of the Wyckoff position of the crystal, the Pb1 ions on the two triangles are referred to as the $6h$ position, while the Pb2 ions on the large hexagon are referred to as the $4f$ position\cite{zhang2023structural}. In the direction of the c axis, the Pb2 ions, together with the surrounding PO$_4$ units, contribute to the formation of a cylindrical column centrally positioned around O2 ions, as that illustrated in Fig.\ref{crystal structure1}(b). It is important to note that the O2 ions are only partially occupied, with a filling of $1/4$.

The calculated electronic band structure and Orbital-resolved density of states (DOS) for Pb$_{10}$(PO$_4$)$_6$O are demonstrated in Fig.\ref{crystal structure1}(c) and Fig.\ref{crystal structure1}(d). Evidently, lead apatite is an indirect band gap semiconductor, with a PBE-predicted indirect band gap of 2.78 eV. This finding confirms well to experimental observed transport properties\cite{lee2023roomtemperature}. As that can be seen in Fig.\ref{crystal structure1}(c), this material is not spin-polarized, so the net magnetic moment of all the atoms is close to zero, and it does not exhibit macroscopic magnetism. It is crucial to emphasize that the $1/4$ occupation of O2 ions plays a vital role in the formation of the band gap. A fully occupied state of O2 ions results in the metallic behavior\cite{Lai_2023}. The Orbital-resolved DOS reveal that the valence bands situated below the Fermi level are primarily contributed by the O$\text{-}2p$ states, while the Pb$\text{-}6p$ states and Pb$\text{-}6s$ states exert a minor influence. The conduction band above the Fermi energy is mainly contributed by the Pb$\text{-}6p$ state. Moreover, two flat bands below the Fermi level can be observed in the A-L-H-A region. Flat bands have weaker dispersion, which means that flat band electrons are more localized. Localized electrons often lead to strong correlation effects in the system, which in turn lead to novel quantum physical phenomenon, such as high-temperature superconductivity, quantum Hall ferromagnetic states, and so on.

\begin{figure}[tbp]
\includegraphics[scale=0.1]{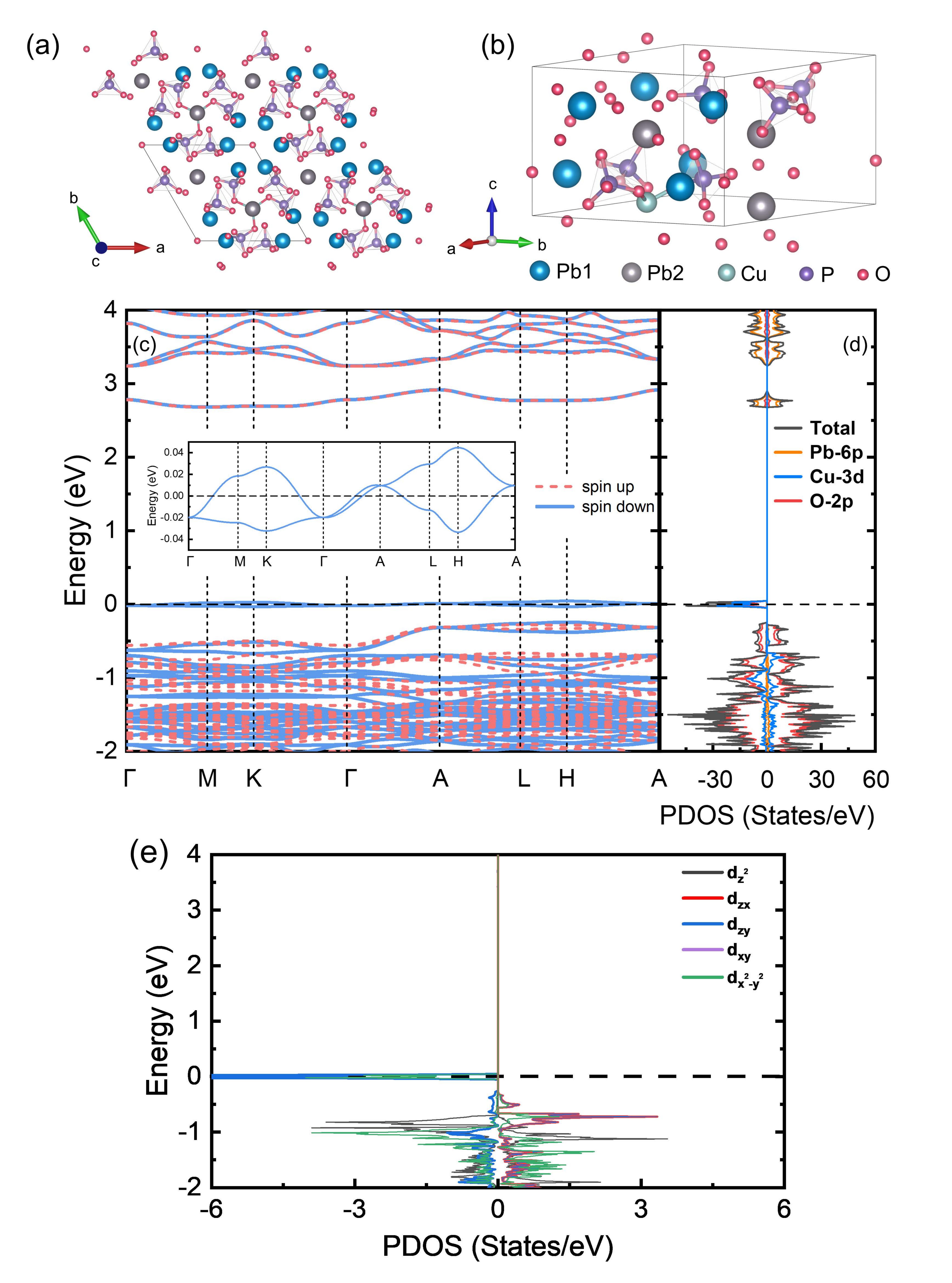}
\caption{(a) Top view of the crystal structure of 4$f$-phase Pb$_9$Cu(PO$_4$)$_6$O compound. The black line denotes the unit cell. (b) Side view of the crystal structure of 4$f$-phase Pb$_9$Cu(PO$_4$)$_6$O compound. (c) The spin polarization band structure of 4$f$-phase Pb$_9$Cu(PO$_4$)$_6$O compound. The Fermi energy level is set to zero. The inset shows the band structure within a small energy window. (d) The spin polarization orbital-resolved DOS of 4$f$-phase Pb$_9$Cu(PO$_4$)$_6$O compound. (e) The spin polarization orbital orientation resolved DOS of 4$f$-phase Pb$_9$Cu(PO$_4$)$_6$O compound.
} \label{crystal structure2}
\end{figure}

Fig.\ref{crystal structure2}(a) and Fig.\ref{crystal structure2}(b) show the crystal structure of 4$f$-phase LK$\text{-}$99. Having analyzed structural and electronic properties of lead-apatite, we turn to 4$f$-phase LK$\text{-}$99. For simplicity, we only consider the case $x=1$, Pb$_9$Cu(PO$_4$)$_6$O. We note that Cu-doping induces symmetry breaking, resulting in four possible configurations. According to the theoretical study by Chen et al., this configuration has the lowest energy when replacing the Pb2 ion with the $4f$ Cu ion, whose crystal structure is shown in Fig.\ref{crystal structure2}(a) and Fig.\ref{crystal structure2}(b)\cite{Lai_2023}. The calculated lattice parameters of 4$f$-phase LK$\text{-}$99 with $(a=9.942 \AA$, $c=7.411 \AA$ and the volume of a unit cell is $620.31 \AA^3$, are in good agreement with the experimental values of $a=9.843 \AA$, $c=7.428 \AA$ and the volume of a unit cell is $623.24 \AA^3$\cite{lee2023superconductor}. As compared to the non-doped lead-apatite, Cu-doping should result in a $0.47\%$ reduction of the system volume and $0.35\%$ shrinkage of crystal constants. 4$f$-phase LK$\text{-}$99 has weakly magnetism whose net magnetic momentum is 1 $\mu_B$ in the unit cell. It is worth mentioning that the magnetic moment of unit cell is $60\%$ provided by Cu ions and the remaining small amount of magnetic moment is distributed on the tetrahedral oxygen composed of Cu, indicating that the doping of Cu ions does greatly affect the properties of lead apatite.

Figs.\ref{crystal structure2}(c-e) show the calculated band structure and DOS of 4$f$-phase LK$\text{-}$99. Due to the presence of spin polarization in this system, we represent results with spin up and spin down by dashed red line and solid blue line in Fig.\ref{crystal structure2}(c), respectively. After Cu-doping, there exist two half-filled flat bands (dispersion$<0.1$eV) crossing the Fermi energy level, and both flat bands are contributed by electrons with spin down. As that can be seen from the embedding diagram, the two flat bands do have two saddle points at the high symmetry points M and L, respectively. This is consistent with the theoretical calculations previously reported by Chen et al\cite{Lai_2023}. Spin-down electrons exhibit the properties of a metal, while spin-up electrons exhibit the properties of an insulator, so the material as a whole show half-metallic property.

In the DOS with spin polarization, the spin-up electron contribution has a positive state density, while the spin-down electron contribution has a negative state density.
As that demonstrated in Fig.\ref{crystal structure2}(d), the bands above the Fermi energy level are fully spin degeneracy, and the bands below the Fermi energy level are partially spin-polarized due to the strong orbital hybridization prior to Cu$\text{-}3d$ and O$\text{-}2p$. The two bands through the Fermi energy level are also mainly contributed by the Cu$\text{-}3d$ and O$\text{-}2p$ orbitals. As a result, conduction electrons are confined to the layer containing Cu$\text{-}$O atoms, while other layers without Cu$\text{-}$O atoms appear to be insulating. In addition, the PO$_4$ units around the cylinders formed by Pb1 ions also exhibit insulating properties, resulting in a one-dimensional conduction channel along the c-axis mediated by the $1/4\text{-}$occupied O2 ions\cite{hao2023firstprinciples}.

The bands near the Fermi energy level are mainly contributed by $d_{zy}$ and $d_{x^2-y^2}$ in the $3d$ orbitals, which are shown in Fig.\ref{crystal structure2}(e).
Since the doping of Cu ions leads to lattice distortion of the parent lead apatite structure, it causes symmetry breaking of the crystal field composed of Cu$\text{-}$O ions, which leads to split of the energy levels of the $3d$ orbitals.

\begin{figure}[tbp]
\includegraphics[scale=0.05]{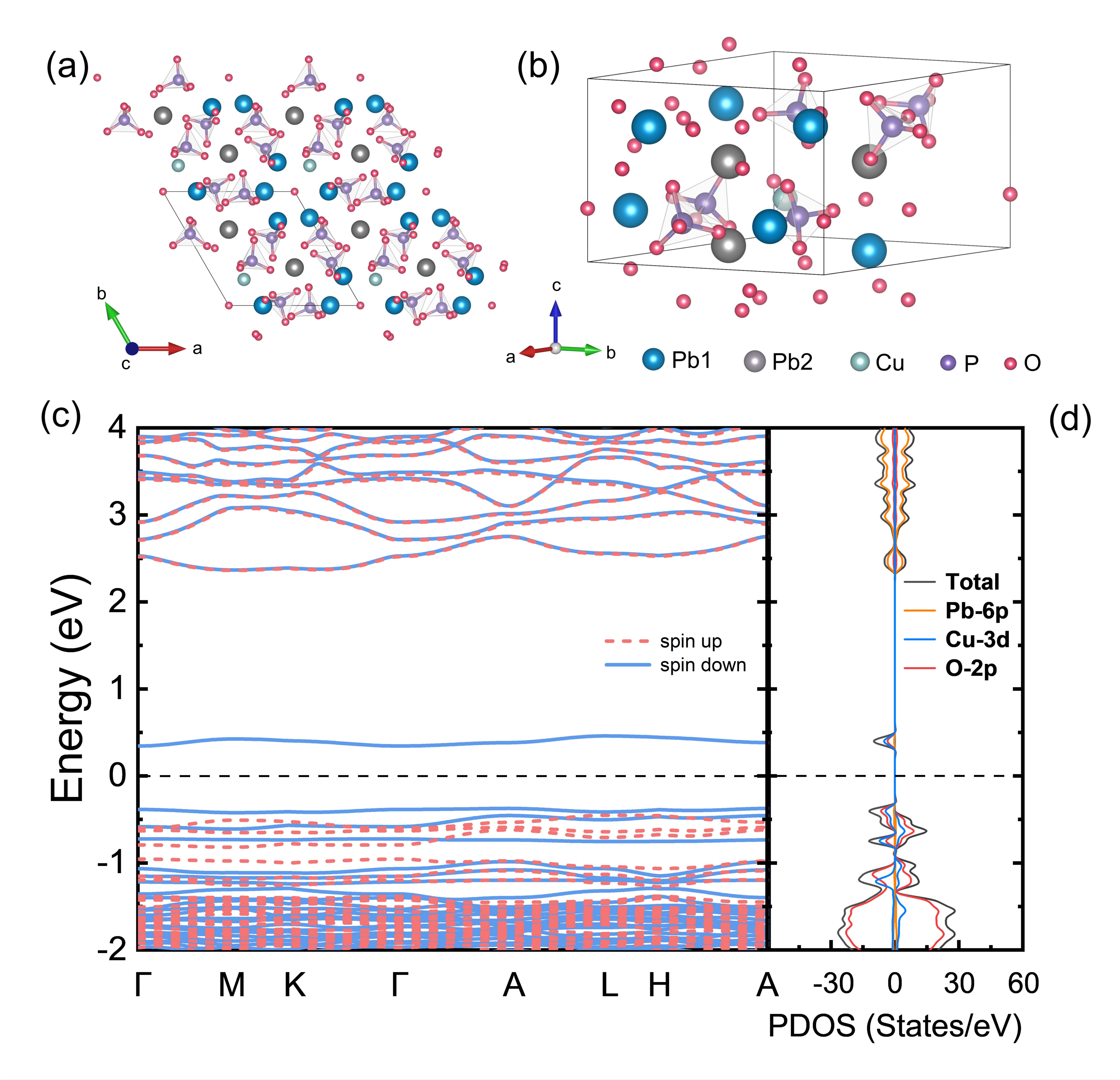}
\caption{(a) Top view of the crystal structure 6$h$-phase of Pb$_9$Cu(PO$_4$)$_6$O compound. The black line denotes the unit cell. (b) Side view of the crystal structure 6$h$ phase of Pb$_9$Cu(PO$_4$)$_6$O compound. (c) The spin polarization band structure of 6$h$-phase of Pb$_9$Cu(PO$_4$)$_6$O compound. The Fermi energy level is set to zero. (d) The spin polarization orbital-resolved DOS of 6$h$-phase of Pb$_9$Cu(PO$_4$)$_6$O compound.
} \label{6hphase}
\end{figure}

Fig. \ref{6hphase}(a) and Fig. \ref{6hphase}(b) illustrate the crystal structure of 6$h$-phase Pb$_9$Cu(PO$_4$)$_6$O.
After performing a full structural optimization, the lattice constants are determined to be $a=9.911\AA, c=7.510 \AA$ and the volume of a unit cell is 638.8 $\AA^3$, which are in good agreement with the experimental values of $a=9.865 \AA, c=7.431 \AA$ and the volume of a unit cell is 626.3 $\AA^3$\cite{lee2023superconductor}. When copper ions are used for doping to replace the Pb1 ion at the 6$h$ position, this leads to a breakdown of lattice symmetry, resulting in six possible conformations. However, at the first step, we will focus on the nature of the electronic structure of the lowest energy configuration. According to the theoretical study by Zhang et al. this configuration has the lowest energy when replacing the Pb1 ion at the 6$h$ position with a Cu ion, and its crystal structure is shown in Fig. \ref{6hphase}(a) and (b).

It is noteworthy that the formation energy of the energy-lowest configuration of the Cu ion replacing the 6$h$ site of Pb ion is even lower than that of the energy-lowest configuration of the Cu ion replacing the 4$f$ site of Pb ion, so it is highly probable that the LK-99 material actually synthesized in  experiments\cite{lee2023superconductor} is the 6$h$-phase LK-99, not the 4$f$-phase LK-99, which had been the subject of more theoretical studies before.
Fig. \ref{6hphase}(c) and (d) show the band structure and DOS for the 6$h$ phase of Pb$_9$Cu(PO$_4$)$_6$O. An isolated impurity energy band exists above the Fermi energy level around 0.45 eV, and the DOS for this band shows that this band is mainly contributed by O atoms, with a conduction band spin-flip band gap of 1.9 eV. An isolated impurity energy band also exists below the Fermi energy level around 0.38 eV, and the closer impurity energy band below the exists obvious spin polarization in the following bands, and the spin-flip band gap in the valence band is 0.07 eV. The overall semiconductor band gap of the material is 0.84 eV, so the material was a bipolar magnetic semiconductor. The carrier spin direction of the bipolar magnetic semiconductor can be directly regulated by simply changing the positive and negative polarity of the applied gate voltage. The material can be utilized as a new type of electric field-modulated spintronics device.

\begin{figure}[tbp]
\includegraphics[scale=0.05]{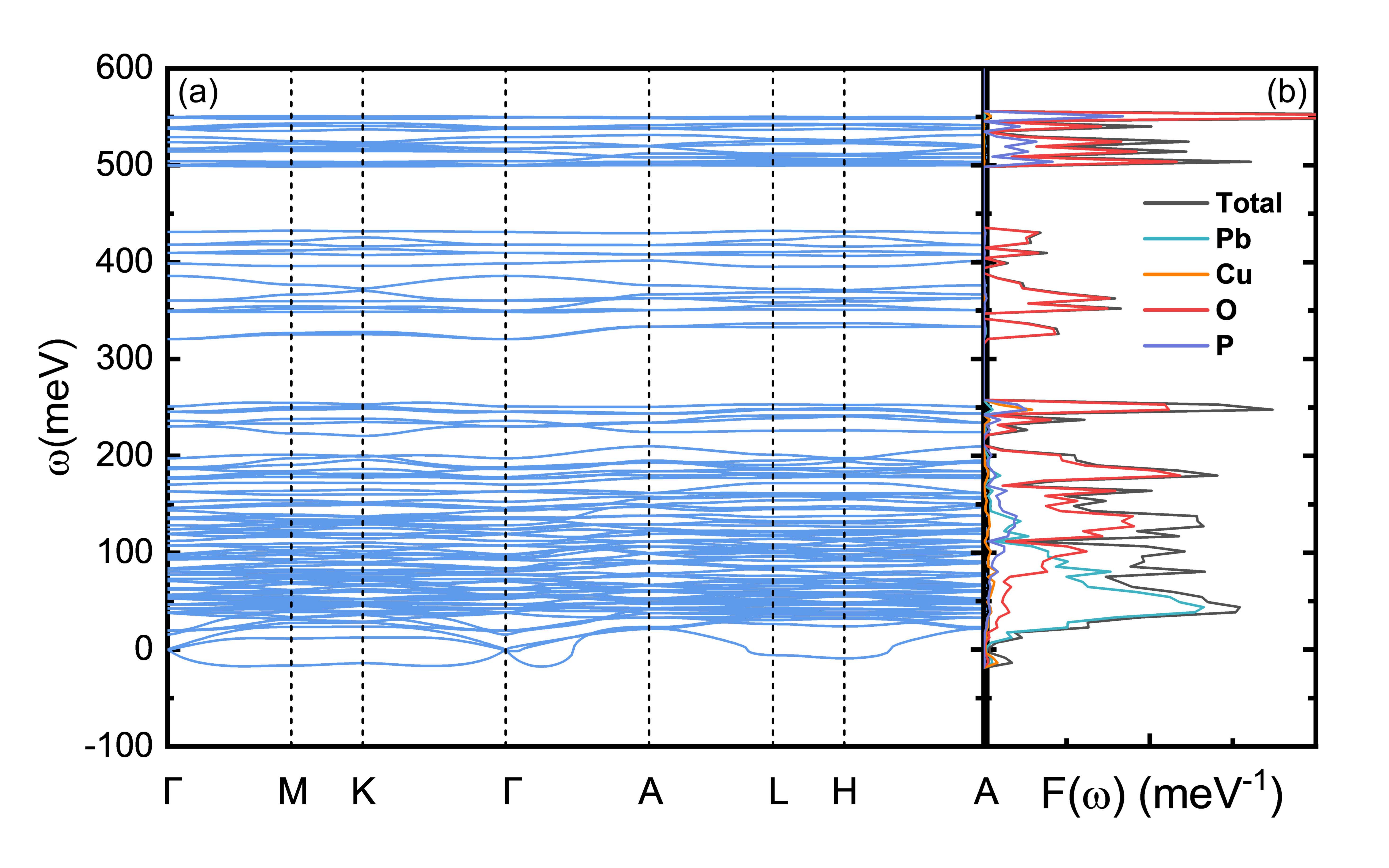}
\caption{(a) Phonon dispersion spectrum and (b) DOS of 4$f$-Pb$_9$Cu(PO$_4$)$_6$O compound.
} \label{phonon}
\end{figure}


The calculated phonon spectra of the 4$f$-phase Pb$_9$Cu(PO$_4$)$_6$O compound is shown in Fig. \ref{phonon}. It can be seen that the phonon spectrum has imaginary frequencies below zero, which proves that the dynamics of the material is unstable, and the phonon density of states shows that a negative acoustic branch of the material has a contribution from the vibration of the Pb atoms, which may be related to the localized lattice distortion induced by the Cu doping, and the detailed micro theoretical explanations need to be further investigated. According to our calculations, the ground state energy of LK-99 in the 6$h$ phase is smaller than that of the 4$f$ phase (the energy difference is 0.10 eV), and since the phonon spectrum of the 4$f$ phase has only small imaginary frequencies, we expect that there is a high probability that the phonon spectrum of the 6$h$ phase is kinetically stabilized, and detailed calculations are needed for further theoretical research.

\noindent
{\it Conclusions} In summary, we have calculated and analyzed the crystal structures, band structures, lattice dynamics, and magnetic properties
of the parent Pb$_{10}$(PO$_4$)$_6$O compounds and two different phases of LK-99 compounds by using the first principles method.
The results of band and DOS calculations indicate that the Pb$_{10}$(PO$_4$)$_6$O compound is an indirect band gap semiconductor. However, the 4$f$-phase LK-99 behaves as a half-metal due to two half-filled flat bands spanning the Fermi energy levels, and further theoretical studies are needed to determine whether it is a superconductor or not. On the contrary, since the 6$h$-phase LK-99 is spin-polarized at the bottom of the conduction band and at the top of the valence band, the material is a bipolar magnetic semiconductor, which explain well the recent large number of experimental studies on the repetition of the experiments in South Korea. Some experiments reports that LK-99 is a superconductor and some reports that it is a semiconductor, which is probably due to the fact that the experimental synthesis of the LK-99 with the Cu ions doped with different sites of Pb ions. In the video of the antimagnetic experiment of LK-99, the material is only semi-suspended and not fully suspended, which is not consistent with the complete diamagnetism of superconductors. Combined with our theoretical study, we speculate that the LK-99 material used in the experimental video may be a mixture of the 4$f$-phase and 6$h$-phase of LK-99.

\noindent
{\emph{ Acknowledgments:}} 
This work is supported by NSCFs (Grant Nos. 11974049).  We also acknowledge the computational support from the HSCC of Beijing Normal University.

\bibliography{ref}
\end{document}